\documentclass{egpubl}
\usepackage{pg2023}

\usepackage[T1]{fontenc}
\usepackage{dfadobe}  
\usepackage{subfiles} % Best loaded last in the preamble
\usepackage{cite}  % comment out for biblatex with backend=biber
\usepackage{kotex}
\usepackage{amsmath}
\usepackage{amsfonts}
\BibtexOrBiblatex
\electronicVersion
\PrintedOrElectronic
\ifpdf \usepackage[pdftex]{graphicx} \pdfcompresslevel=9
\else \usepackage[dvips]{graphicx} \fi

\usepackage{egweblnk} 
\title[DeepIron]%
      {DeepIron: Predicting Unwarped Garment Texture from a Single Image}

\author[Hyun-Song Kwon, Sung-Hee Lee]
{\parbox{\textwidth}{ \centering Hyun-Song Kwon$^{1}$, Sung-Hee Lee$^{1}$
        }
        \\
{\parbox{\textwidth}{\centering $^1$Korea Advanced Institute of Science and Technology, Republic of Korea%
       }
}
}
\begin{document}

% uncomment for using teaser
 \teaser{
  \includegraphics[width=\linewidth]{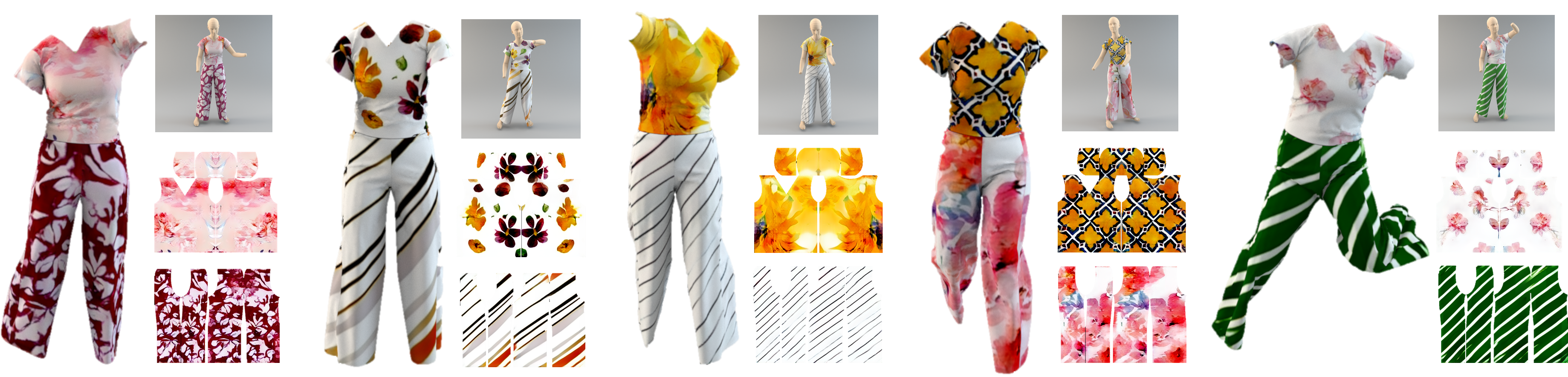}
  \centering
   \caption{We introduce DeepIron - a framework for reconstructing 3D garments by inferring the unwarped original texture of the input garment. The inferred unwarped textured images allows to create realistic appearance of 3D garments when deformed to fit new poses.
}}
 \label{fig:teaser}
%}

\maketitle
%-------------------------------------------------------------------------
\begin{abstract}
Realistic reconstruction of 3D clothing from an image has wide applications, such as avatar creation and virtual try-on. This paper presents a novel framework that reconstructs the texture map for 3D garments from a single image with pose. Assuming that 3D garments are modeled by stitching 2D garment sewing patterns, our specific goal is to generate a texture image for the sewing patterns. A key component of our framework, the Texture Unwarper, infers the original texture image from the input clothing image, which exhibits warping and occlusion of texture due to the user's body shape and pose. The Texture Unwarper effectively transforms between the input and output images by mapping the latent spaces of the two images. By inferring the unwarped original texture of the input garment, our method helps reconstruct 3D garment models that can show high-quality texture images realistically deformed for new poses. We validate the effectiveness of our approach through a comparison with other methods and ablation studies. 

%The predicted texture fill inside the sewing patterns which correspond to the clothing in the input.
%, we can not only represent the 3D garment mesh with natural shapes, and detailed wrinkles in any pose but also further distort the texture to match that specific pose.
%By releasing this dataset, we expect to be able to help many researchers in 3D garment-related field.
%Our research presents a novel methodology for rectifying distorted garment textures, offering a fresh perspective in the field. As a result, we anticipate substantial advancements in the development of virtual try-on systems and the generation of virtual human datasets. 

%-------------------------------------------------------------------------
%  ACM CCS 1998
%  (see https://www.acm.org/publications/computing-classification-system/1998)
% \begin{classification} % according to https://www.acm.org/publications/computing-classification-system/1998
% \CCScat{Computer Graphics}{I.3.3}{Picture/Image Generation}{Line and curve generation}
% \end{classification}
%-------------------------------------------------------------------------
%  ACM CCS 2012
   %(see https://www.acm.org/publications/class-2012)
%The tool at \url{http://dl.acm.org/ccs.cfm} can be used to generate
% CCS codes.
%Example:


\end{abstract}  
%-------------------------------------------------------------------------
\section{Introduction}
\documentclass[../pg2023main.tex]{subfiles}
\begin{document}

The acquisition of high-quality 3D garment models is becoming increasingly important for creating digital humans in various fields, such as feature films, virtual reality, and digital fashion. 

Researchers have developed techniques to reconstruct 3D garment models from 2D images or 3D scan data \cite{ patel2020tailornet, moon20223d, jiang2020bcnet, he2021arch++, huang2020arch,  mir2020learning, majithia2022robust, xu20193d, zakharkin2021point}. 
However, most of these techniques have primarily focused on reconstructing detailed geometry, neglecting the extraction of high-quality garment texture, which often leads to producing blurry textures.

Reconstructing garment textures from images presents a challenging task because garments are heavily deformed by the body shape and pose, and are occluded by other body parts and wrinkles. 
Some studies have made advancements in restoring garment textures from input images \cite{mir2020learning, majithia2022robust,xu20193d}. These studies enhanced texture quality by finding the area in the input image that corresponds to the garment and mapping it to the UV map.

However, their methods have limitations in predicting textures for occluded parts of the garment.
Furthermore, when the garment undergoes deformation and develops wrinkles due to changes in pose, these methods produce unnatural texture images. 
This is because these studies primarily focus on accurately reconstructing the particular pose in the input rather than attempting to predict the original texture before the distortion occurs.

To address this limitation, we approach the problem by using garment sewing patterns as the representation of the garment shape and inferring the texture image for the sewing patterns. As garments are constructed by stitching the sewing patterns, reconstructing 3D garments in terms of their fundamental sewing patterns and the associated texture image is the most principled way. 
Stitching the sewing patterns and draping them over the body using physical simulation results in natural deformation of garments including wrinkles for various poses.

In this paper, we propose a novel framework to automatically generate texture maps in the form of garment sewing patterns filled with distortion-corrected texture image from input images. 
Central to our framework is the Texture Unwarper, which effectively transforms the distorted and occluded garment image to its original texture image through an intermediate module that maps the latent space of the input and output images.
Our strategy of separately training each module of the Texture Unwarper increases the quality of output texture image by effectively training each module.
The resulting distortion-corrected image from the Texture Unwarper is then converted into a texture map for the entire garment, ready to be simulated to construct 3D garment in subsequent steps.

In this work, we focus on inferring texture image robust to pose variation. Specifically, we set a female body as our test subject as it includes more challenging curves than male body. Other variations in terms of body shape and garment sewing pattern are not considered by using only one body shape (a female SMPL model) and one sewing pattern for each type of garments, including T-shirt and pants.

In summary, our contributions are as follows:
\begin{itemize}
\item We propose the Texture Unwarper, which corrects distorted garment textures associated with the deformed garment in input images. By generating un-distorted original texture image, we can produce natural garment appearance when deformed to match new poses.
% \item We release a dataset including garment texture maps of over 10K samples and 50K synthetic images of a female body wearing garments with various poses.
\end{itemize}
\end{document}